\documentclass{article}

\def\abstract#1{\vskip 7mm 
        \begin{center}{\large Abstract}\par \smallskip
                \begin{minipage}[c]{12cm}
                        \small #1
                \end{minipage}
        \end{center}
}
\def\title#1{\begin{center}{\Large\bf #1}\end{center}}
\def\author#1{\vskip 5mm \begin{center}{#1}\end{center}}
\def\address#1{\begin{center}{\it #1}\end{center}}

\usepackage{amsmath}
\usepackage[mathscr]{eucal}
\usepackage[dvips]{graphicx}
\usepackage{psfrag}

\newcommand{\del}[1]{\partial_{#1}}

\newcommand{\vect}[1]{\ensuremath{\boldsymbol{#1}}}

\psfrag{k}{$\vect{k}$}
\psfrag{q}{$\vect{q}$}
\psfrag{kq}{$\vect{k}-\vect{q}$}
\psfrag{a}{the propagator of the scalar field$\,\psi\,$}
\psfrag{b}{the propagator of the vector field$\,\vect{A}\,$}

\psfrag{k1}{$\vect{k}$}
\psfrag{q1}{$\vect{q}$}
\psfrag{kq1}{$\vect{k}+\vect{q}$}

\psfrag{psi}{$|\psi|^2$}
\psfrag{r}{$\tilde{r}$}

\makeatletter
\@ifundefined{lesssim}{}{}
\@ifundefined{gtrsim}{}{}
\def\vereq#1#2{\lower3pt\vbox{\baselineskip1.5pt \lineskip1.5pt
\ialign{$\m@th#1\hfill##\hfil$\crcr#2\crcr\sim\crcr}}}
\makeatother

\begin{document}

\title{Finite Temperature Field Theory of ``Extreme Black Holes''}
\author{%
  Yoshitaka Degura\footnote{E-mail:c1997@sty.sv.cc.yamaguchi-u.ac.jp} and %
  Kiyoshi Shiraishi\footnote{E-mail:shiraish@sci.yamaguchi-u.ac.jp}
}
\address{%
  Graduate School of Science and Engineering, Yamaguchi University, \\
  Yoshida, Yamaguchi-shi, Yamaguchi 753--8512, Japan
}

\abstract{%
We treat the model which describes ``extreme black holes" moving slowly. We derive an effective lagrangian in the low energy for this model and then investigate a statistical behavior of ``extreme black holes" in the finite temperature.
}

\section{Introduction}

``Extreme black holes" are the objects (solitons) which have a certain condition between electric charge and mass; these two horizons coincide with each other. In other words, ``extreme black holes" realize when three forces (the gravitational force, the electric force and the dilaton force) between dilatonic black holes are cancelled with each other. We regard ``extreme black holes"  as point particles, so we consider that the collection of these particles is expressed by the scalar field. For the static case, we know that the solution of ``extreme black holes" is the Papapetrou-Majumdar-Myers solution \cite{ref1, ref2, ref3}. On the other hand, we consider ``extreme black holes" for the dynamical case, we derive a theoretical model of a (non-relastivistic) field including the low energy interaction  of ``extreme black holes". In the finite temperature, we examine the structure of the gas of ``extreme black holes". As a result, we will show that the gas of ``extreme black holes" lumps at the high temperature.

\section{Effective Lagrangian}

In this section, we derive an effective lagrangian for ``extreme black holes" of which a typical velocity $v$ is slow. In the classical theory, the action of particles, which have mass $m$ and charge $e$, coupled to the gravitational field, the electromagnetic field and the dilaton field, is
\begin{equation}
\label{eq:201}
I = \,- \int\!\! ds\:\Bigl[m\,e^{a\phi} + e A_\mu \frac{dx^\mu}{ds} \Bigl]\, ,
\end{equation} 
where $\phi$ is a dilaton field,\, $a$ is the coupling constant of the dilaton field.
Then, the four-momentum of the particle is
\begin{equation}
\label{eq:202}
 P_{\mu} = m e^{a\phi} g_{\mu\nu} \frac{dx^{\nu}}{ds} - e A_{\mu}.
\end{equation}
This four-momentum satisfies the following equation:
\begin{equation}
\label{eq:203}
 g^{\mu\nu}(P_{\mu} + e A_{\mu})(P_{\nu} + e A_{\nu}) + m^2 e^{2a\phi} = 0 .
\end{equation}

Next, in quantum theory, we can rewrite this equation (\ref{eq:203}) into a wave equation for a wave function $\varphi$:
\begin{equation}
\label{eq:204}
 \Bigl[\,g^{\mu\nu}(P_{\mu} + e A_{\mu})(P_{\nu} + e A_{\nu})  %
          + m^2 e^{2a\phi} \,\Bigr]\,\varphi = 0  .
\end{equation} 
So the action which gives this wave equation is
\begin{eqnarray}
\label{eq:205}
 S_m\!&=&\!\int\!\!d^4 x\,\sqrt{-g}\:\Bigl[-\,\varphi^{*}e^{-a\phi}g^{\mu\nu} %
               (P_{\mu} + e A_{\mu})(P_{\nu} + e A_{\nu})\varphi  %
        - m^2 e^{a\phi} \varphi^{*}\varphi \Bigl] ,
\end{eqnarray}  
where we regard $\varphi$ as a ``field" from now on. Therefore the total action including interactions is
\begin{eqnarray}
\label{eq:206}
 S &=& \int\!\! d^4 x\,\frac{\sqrt{-g}}{16\pi}\:\Bigl[\,R-2(\nabla_{\mu}\phi)%
           (\nabla^{\mu}\phi) - e^{-2a\phi} F_{\mu\nu}F^{\mu\nu}\Bigl] %
          \:+ \;S_m ,
\end{eqnarray}  
which leads to the following field equations,
 \begin{eqnarray}
\label{eq:207} 
 &&  \nabla^2 \phi + \frac{a}{2}\, e^{-2a\phi} F^2  %
   +\,4\pi a \Bigl[\,e^{-a\phi} \,\varphi^{*} (P + e A)^2 \varphi %
  - e^{a\phi} m^2 \,\varphi^{*} \varphi\, \Bigr] = 0 ,\\
\notag && \\
\notag 
 &&R_{\mu\nu} - \frac{1}{2}\,g_{\mu\nu}R %
   = 2\,\Bigl[\,\nabla_{\mu}\phi\nabla_{\nu}\phi%
       -\frac{1}{2}\,g_{\mu\nu}(\nabla\phi)^2 \Bigr]  %
 + \:e^{-2a\phi}\Bigl[\,2\,F_{\mu\nu}^2 - \frac{1}{2}\,g_{\mu\nu}F^2 \Bigr] \\
 &&\notag \qquad\qquad\qquad\qquad
  + 16\pi\biggl\{e^{-a\phi} Re\Bigl[\varphi^{*}(P_{\mu} + e A_{\mu}) %
            (P_{\nu} + e A_{\nu})\,\varphi \\  
 && \label{eq:208} \qquad\qquad\qquad\qquad\qquad\qquad
 - \frac{1}{2}\,g_{\mu\nu}\varphi^{*}(P + e A)^2 \varphi\Bigr] %
    - \frac{1}{2}\,g_{\mu\nu} e^{a\phi} m^2 \varphi^{*} \varphi\, \biggr\} , \\
\notag && \\
&&  \nabla_{\mu}\Bigl[\,e^{-2a\phi} F^{\mu\nu} \Bigr] = 8\pi e\,e^{-a\phi} %
             \varphi^{*}g^{\nu\lambda}(P_{\lambda} + e A_{\lambda})\,\varphi .
 \label{eq:209}
 \end{eqnarray}
 
From now on we consider in the lowest order of a typical velocity $v$ of ``extreme black holes", so we assume as follows:
\begin{eqnarray}
\label{eq:210}
 ds^2 &=& - U^{-2}\bigl(dt + B_i dx^i \bigr)^2 + U^2 d \vect{x}^2 \:\: , \\
\label{eq:211}
 U &=& V(\vect{x})^{\frac{1}{1+a^2}} \:\: , \\
\label{eq:212}
 e^{-2a\phi} &=& V^{\frac{2 a^2}{1+a^2}} \:\: , \\
\label{eq:213}
 A_0 &=& \frac{1}{\sqrt{1+a^2}}\Bigl(1 -\frac{1}{V}\Bigr)  \:\: , \\
\label{eq:214}
 A_i \sim B_i  &=&  O(v) .
\end{eqnarray} 
Under these ansatze, for a static configuration ($A_i = B_i = 0$), a vacuum solution satisfies
\begin{equation}
\label{eq:215}
 \partial^2 V = 0 ,
\end{equation}
which means that this vacuum solution represents an arbitrary number of ``extreme black holes" \cite{ref1, ref2, ref3}. One can find that the relation between mass $m$ and electric charge $e$ of a particle is, as seen from the correspondence of ``extreme black holes",
\begin{equation}
\label{eq:216}
 \frac{e}{m} = \sqrt{1 + a^2}.
\end{equation}

Next, we consider the low energy limit, $ - \,P_0 - m = E - m \ll m$.
Then
\begin{eqnarray}
\notag
 P_0 + e A_0 &=& P_0 + \frac{e}{\sqrt{1+a^2}}\,\Bigl(1-\frac{1}{V}\Bigr) %
                           = P_0 + \Bigl(1-\frac{1}{V}\Bigr)  \\
             &\approx& - \,m \,\frac{1}{V} ,
\label{eq:218} \\
\notag && \\
\notag
 P_i + e A_i - B_i(P_0 + e A_0) &\approx& P_i + %
       e\,\Bigl(A_i + \frac{1}{\sqrt{1+a^2}}\,\frac{1}{V}\,B_i\Bigr)  \\
             &\equiv& P_i + e \hat{A}_i ,  \label{eq:219}
\end{eqnarray}
where
\begin{equation}
\label{eq:220}
 \hat{A}_i \equiv A_i + \frac{1}{\sqrt{1+a^2}}\,\frac{1}{V}\,B_i \, ,
\end{equation}
and
\begin{eqnarray}
\notag
\hat{F}_{ij}  &\equiv& \del{i}\hat{A}_j - \del{j}\hat{A}_i \\
             &=& \tilde{F}_{ij} - \frac{1}{\sqrt{1+a^2}}\,\frac{1}{V}\,G_{ij},
\label{eq:221}
\end{eqnarray}
where
\begin{eqnarray}
\label{eq:222}
\tilde{F}_{ij}  &\equiv& F_{ij} + B_i F_{j0} - B_j F_{i0} , \\
\label{eq:223}
G_{ij}          &\equiv& \del{i}B_j - \del{j}B_i.
\end{eqnarray}

Taking the low energy and non-relativistic limit %
$- P_0 - m = E - m \ll m$, $|P_i + e\hat{A}_i|^2 \approx m^2 v^2 \ll m^2$,
we solve explicitly the field equations (\ref{eq:207}), (\ref{eq:208}) and (\ref{eq:209}). From the dilaton field equation (\ref{eq:207}), the ($0\;0$) component of the gravitational field equation (\ref{eq:208}) and the time component of the electromagnetic field equation (\ref{eq:209}), in the lowest order, we obtain
\begin{equation}
\label{eq:225}
\partial^2 V + 8\pi (1 + a^2) m^2 U^3 |\varphi|^2 = 0.
\end{equation}
From the ($0\; i$) component of the gravitational field equation (\ref{eq:208}), we get
\begin{eqnarray}
\notag
\del{\ell}\Bigl[V^{\frac{2(a^2-1)}{1+a^2}}\,\frac{1}{V^2}\,G_{\ell i}\Bigr]&=&%
    4\,\sqrt{\frac{1}{1+a^2}}\,\del{\ell}\Bigl[V^{\frac{2(a^2-1)}{1+a^2}}\,%
                                \frac{1}{V}\,\tilde{F}_{\ell i}\Bigr]  %
    -\,4\,\sqrt{\frac{1}{1+a^2}}\,\frac{1}{V}\,\del{\ell}\Bigl[%
           V^{\frac{2(a^2-1)}{1+a^2}}\,\tilde{F}_{\ell i}\Bigr]  \\
  &&\quad
   + \:32\pi\,\frac{m}{V}\;e^{-a\phi}\,\varphi^{*}(P_{i} + e A_{i})\,\varphi ,
  \label{eq:226}
\end{eqnarray}
On the other hand, the space component of the electromagnetic field equation (\ref{eq:209}) reads
\begin{equation}
\label{eq:227}
\del{\ell}\Bigl[V^{\frac{2(a^2-1)}{1+a^2}}\,\tilde{F}_{\ell i}\Bigr] %
= 8\pi e\,e^{-a\phi}\,\varphi^{*}(P_{i} + e A_{i})\,\varphi .
\end{equation}
 We define an antisymmetric tensor field $H_{ij}$ as
\begin{equation}
\label{eq:228}
H_{ij} \equiv 4\,\sqrt{\frac{1}{1+a^2}}\tilde{F}_{ij} - \frac{1}{V}\,G_{ij}.
\end{equation} 
From eqs.~(\ref{eq:226}) and (\ref{eq:227}), the antisymmetric tensor field equation is
\begin{equation}
\label{eq:229}
\del{\ell}\Bigl[V^{\frac{2(a^2-1)}{1+a^2}}\,\frac{1}{V}\,H_{\ell i}\Bigr] = 0.
\end{equation} 
As stated, the ($0\; i$) component of the gravitational field equation (\ref{eq:208}) and the space component of the electromagnetic field equation (\ref{eq:209}) are
\begin{equation}
\label{eq:231}
-\,\frac{1+a^2}{3-a^2}\,\del{\ell}\Bigl[V^{\frac{2(a^2-1)}{1+a^2}}\,\hat{F}_%
{\ell i}\Bigr] = 8\pi e\,e^{-a\phi}\,\varphi^{*}(P_{i}+e\hat{A}_{i})\,\varphi ,
\end{equation}
By the way, from eqs.~(\ref{eq:221}) and (\ref{eq:228}), we can obtain the following relation:
\begin{equation}
\label{eq:233}
\frac{1}{4}\,\frac{1}{V^2}\,G^2 - \tilde{F}^2 = %
             \frac{1+a^2}{3-a^2}\,\Bigl(\hat{F}^2 - \frac{1}{4}\,H^2\Bigr) .
\end{equation} 

Now we consider the effective lagrangian. From the total action (\ref{eq:206}),
\begin{eqnarray}
\notag
\mathscr{L} &=& \sqrt{-g}\,\biggl\{\frac{1}{16\pi}\:\Bigl[\,R  %
                          - 2\,(\nabla\phi)^2 - e^{-2a\phi}\,F^2\Bigl] \\
\notag &&\qquad\qquad\quad
 + \:\Bigl[- \,\varphi^{*} e^{-a\phi} g^{\mu\nu}(P_{\mu} + e A_{\mu})(P_{\nu} %
   + e A_{\nu})\,\varphi %
   - m^2 e^{a\phi} \varphi^{*}\varphi \Bigl] \biggr\} \\
\notag &&   \\
\notag &\approx& 
  \frac{1}{16\pi}\,V^{\frac{2(a^2-1)}{1+a^2}}\,%
    \biggl[\frac{1}{4}\,\frac{1}{V^2}\,G^2 - \tilde{F}^2 \biggr] \\
\notag && \:
 + \:\Bigl[V\,U^3\varphi^{*}\bigl(P_{0} + e A_{0}\bigr)^2\varphi %
             - \frac{1}{V}m^2 U^3 \varphi^{*}\varphi %
 - \frac{1}{V}\,V^{\frac{2(a^2-1)}{1+a^2}}\,U^3 \varphi^{*}%
                     \bigl(P_{i} + e \hat{A}_{i}\bigr)^2\varphi \Bigr] \\
\notag &&   \\
\notag &=& 
 \frac{1}{16\pi}\,V^{\frac{2(a^2-1)}{1+a^2}}\,%
 \biggl[\frac{1+a^2}{3-a^2}\,\Bigl(\hat{F}^2 - \frac{1}{4}\,H^2\Bigr)\biggr] \\
\label{eq:235} && \:
 + \:\Bigl[VU^3\varphi^{*}\bigl(P_{0} + e A_{0}\bigr)^2\varphi %
             - \frac{1}{V}m^2 U^3 \varphi^{*}\varphi %
 - \frac{1}{V}\,V^{\frac{2(a^2-1)}{1+a^2}}\,U^3 \varphi^{*}%
                     \bigl(P_{i} + e \hat{A}_{i}\bigr)^2\varphi \Bigr] .
\end{eqnarray} 
In the first term of this effective lagrangian, $H_{ij}$ can be regarded as an external field because $H_{ij}$ does not couple to the scalar field $\varphi$. Because there is no external field in the model which we have considered, we consider only the interactions among black holes, thus $H_{ij} \equiv 0$.

We introduce a non-relativistic field $\psi$:
\begin{equation}
\label{eq:236}
\psi \equiv \sqrt{2m}\,U^{3/2}\,\varphi,
\end{equation} \\
where since $(-g^{(3)})^{1/4} = U^{3/2}$, we obtain a correct measure. 
So the effective lagrangian in the low energy is
\begin{eqnarray}
\notag
\mathscr{L} &=& 
   \psi^{*}\,\bigl(-\,P_{0} - m\bigr)\psi - %
            \frac{1}{2m\,V^{(3-a^2)/(1+a^2)}}\,\psi^{*}\,\bigl(\vect{P} %
                        + e \vect{\hat{A}}\bigr)^2\psi \\
&& \qquad + \frac{1}{16\pi}\,\frac{1+a^2}{3-a^2}\,%
                    \frac{1}{V^{2(1-a^2)/(1+a^2)}}\,\hat{F}^2 ,
\label{eq:238}
\end{eqnarray}  
where $V$ satisfies the following equation:
\begin{equation}
\label{eq:239}
\partial^2 V + 4\pi\,(1 + a^2)\,m\,|\psi|^2 = 0 .
\end{equation} 
As a check, varying this effective lagrangian (\ref{eq:238}) with respect to $\hat{\vect{A}}$, we can derive again the field equation (\ref{eq:231}) in the low energy approximation:
\begin{equation}
\label{eq:240}
-\,\frac{1+a^2}{3-a^2}\,\del{\ell}\Bigl[V^{\frac{2(a^2-1)}{1+a^2}}\,\hat{F}_%
{\ell i}\Bigr]=8\pi e\,e^{-a\phi}\,\varphi^{*}(P_{i}+e \hat{A}_{i})\,\varphi .
\end{equation}


\section{Hartree\,-\,Fock Approximation in the Finite Temperature}

First we rewrite the effective lagrangian (\ref{eq:238}) derived in the previous section as follows:
\begin{equation}
\label{eq:241}
\mathscr{L} =  
 \psi^{*}\,i\frac{\partial}{\partial t}\,\psi - \frac{1}{2M}\,\psi^{*}\, %
 \bigl(\vect{P}+\tilde{e} \vect{A}\bigr)^2 \psi+\frac{1}{16\pi}\,F_{ij}^2\, ,
\end{equation}
where we omitted hat for simplicity. We have also used the notations:
\begin{eqnarray}
\label{eq:242}
     M       &\equiv& m\, V^{\frac{3-a^2}{1+a^2}} , \\
\notag 
  \tilde{e}^2  &\equiv& \frac{3-a^2}{1+a^2}\,V^{\frac{2(1-a^2)}{1+a^2}} e^2 \\
\label{eq:243} 
              &=& (3-a^2) V^{\frac{2(1-a^2)}{1+a^2}}\, m^2 .
\end{eqnarray}
The relation between mass $m$ and electric charge $e$ is again
\begin{equation}
\label{eq:244}
 \frac{e}{m} = \sqrt{1 + a^2} .
\end{equation}

Next we consider the field theory in the finite temperature \cite{ref4}. We define the propagator of the vector field $\vect{A}$:
\begin{equation}
\label{eq:245}
\nu_{ij}(\vect{q})=\nu(\vect{q})\Bigl(\delta_{ij}-\frac{q_i q_j}{q^2}\Bigr) .
\end{equation}
In the lowest order, the self-energy of the scalar field is
\begin{figure}[t]
\centering
\psfrag{k}{$\vect{k}$}
\psfrag{q}{$\vect{q}$}
\psfrag{kq}{$\vect{k}-\vect{q}$}
\psfrag{a}{the propagator of the scalar field$\,\psi\,$}
\psfrag{b}{the propagator of the vector field$\,\vect{A}\,$}
\includegraphics*[width=10cm,height=2.5cm]{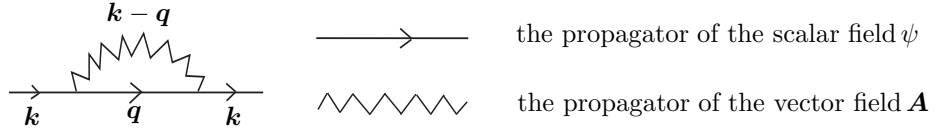}
\caption{Self-energy of the scalar field.}
\label{fig:scalar}
\end{figure}
\begin{eqnarray}
 \Sigma(\vect{k}) &=& -\frac{1}{\beta\mathscr{V}}\frac{1}{4M^2}\sum_{q,\ell} %
          \frac{\nu(\vect{k}-\vect{q})}{i\omega_\ell - \epsilon'_q} %
        \Bigl[(\vect{k}+\vect{q})^2 -\frac{\{(\vect{k}+\vect{q})\cdot %
               (\vect{k}-\vect{q})\}^2}{(\vect{k}-\vect{q})^2}\Bigr] ,
\label{eq:246} 
\end{eqnarray}
(see Fig.~\ref{fig:scalar}) where $\mathscr{V}$ is the volume of the system, $\beta = 1/T$ ($T$ is the temperature of the system), $\omega_{\ell}(\:=\:2\ell\pi/\beta)$ is a frequency and
\begin{eqnarray}
\label{eq:247} 
 \epsilon'_k &=&  \frac{1}{2M}\vect{k}^2 + \Sigma(\vect{k}) -\mu. \\
\notag
             &&  \qquad (\mu\, :\text{a chemical potential.}) 
\end{eqnarray}
In the lowest order, $\nu(\vect{q}) = \tilde{e}^2/\vect{q}^2$. %
We transform the sum over $\vect{q}$, $\ell$  into integral representations;
we can rewrite the self-energy of the scalar field (\ref{eq:246}) as
\begin{eqnarray}
\notag 
 \Sigma(\vect{k}) &=& \frac{\tilde{e}^2}{4M^2}\frac{1}{(2\pi)^3}\int\!    %
            d^3\vect{q} \,f_q \frac{1}{(\vect{k}-\vect{q})^2}        %
 \Bigl[(\vect{k}+\vect{q})^2 - \frac{\{(\vect{k}+\vect{q})\cdot %
                  (\vect{k}-\vect{q})\}^2}{(\vect{k}-\vect{q})^2}\Bigr]   \\
 &=&
          \frac{\tilde{e}^2}{M^2}\frac{1}{4\pi^2}\int_0^\infty\!\!    %
          dq \,q^2 f_q \Biggl[-1-\frac{k^2+q^2}{2kq} %
                         \ln \left|\frac{k-q}{k+q}\right|\Biggr]  ,
\label{eq:251} 
\end{eqnarray}
where $f_q$ is a distribution function of the scalar field:
\begin{eqnarray}
\label{eq:252} 
 f_q &\equiv& \frac{1}{e^{\beta \epsilon'_q} - 1} .
\end{eqnarray}
The self-energy of the scalar field $\Sigma(\vect{k})$ is, by non-relativistic treatment,
\begin{eqnarray}
\label{eq:253} 
 -1-\frac{k^2+q^2}{2kq}\ln |\frac{k-q}{k+q}| \approx \;\frac{4k^2}{3q^2} .
\end{eqnarray}
From eq.~(\ref{eq:253}), there is no constant term in terms of $k$, so can write approximately
\begin{eqnarray}
\label{eq:254} 
 \Sigma(\vect{k}) \:\approx \:\frac{1}{2}\,B k^2 ,
\end{eqnarray} 
where $B$ is a constant.

From now on we consider the high temperature approximation. Then the distribution function of the scalar field is
\begin{eqnarray}
\label{eq:255} 
 f_q &\approx& \;e^{-\beta \epsilon'_q} 
\end{eqnarray}
Therefore (self-consistent) Hartree\,-\,Fock equation is 
\begin{eqnarray}
\notag
 \Sigma(\vect{k}) &\approx& \;\frac{1}{2}\,B k^2 \\
 \notag           &\approx& 
\frac{\tilde{e}^2}{4 \pi^2M^2}\int_0^\infty\!dq\,q^2 f_q\,\frac{4k^2}{3q^2} \\
  &=&
\frac{1}{2}\:\frac{4\,\tilde{e}^2\,e^{\beta\mu}}{3(2\pi)^{3/2}M^2}\,  %
 \Bigl[\beta\bigl(\frac{1}{M} + B\bigr)\Bigr]^{-1/2}\,k^2 .
 \label{eq:256} 
\end{eqnarray} 
As a result, we get
\begin{eqnarray}
\label{eq:257} 
 B = \frac{4\,\tilde{e}^2\,e^{\beta\mu}}{3(2\pi)^{3/2}M^2}\, %
                    \Bigl[\beta\bigl(\frac{1}{M} + B\bigr)\Bigr]^{-1/2} .
\end{eqnarray} 
So the solution of eq.~(\ref{eq:257}) in terms of $B$ is 
\begin{eqnarray}
\label{eq:258} 
 B = \frac{4\,\tilde{e}^2\,e^{\beta\mu}}{3(2\pi)^{3/2}M^2}\,   %
           \sqrt{\frac{M}{\beta}}\Bigl(\cos \frac{\theta}{3}+   %
               \frac{1}{\sqrt{3}}\sin\frac{\theta}{3}\Bigr)^{-1} ,
\end{eqnarray} 
where
\begin{eqnarray}
\label{eq:259} 
 \theta \:= \:\sin^{-1} %
    \frac{2\sqrt{3}\,\tilde{e}^2 e^{\beta\mu}}{(2\pi)^{3/2}\sqrt{\beta M}} .
\end{eqnarray}
On the other hand, the particle density $n$ is
\begin{eqnarray}
\notag
 n &=& \frac{1}{(2\pi)^3}\int\! d^3\vect{q} \,f_q \\
  &=&     \frac{e^{\beta\mu}}{(2\pi)^{3/2}}\, %
                 \Bigl[\beta\bigl(\frac{1}{M}+B\bigr)\bigr]^{-3/2} .
\label{eq:260} 
\end{eqnarray}
The particle density can also be written by $|\psi|^2$, therefore
\begin{eqnarray}
\label{eq:261}
 |\psi|^2= n = %
    e^{\beta\mu}\Bigl(\frac{M}{2\pi\beta}\Bigr)^{3/2}\,\Bigl[%
    \cos\frac{\theta}{3} + \frac{1}{\sqrt{3}}\sin\frac{\theta}{3}\Bigr]^{-3} .
\end{eqnarray}
Now we have to solve the following equation with this equation (\ref{eq:261})
\begin{eqnarray}
\label{eq:263}
\partial^2 V + 4\pi\,(1 + a^2)\,m\,|\psi|^2 = 0 .
\end{eqnarray}

\begin{figure}[t]
\qquad\qquad$(1)\, :\,a^2=0$
\centering
\psfrag{psi}[]{$\tiny{|\psi|^2}$}
\psfrag{r}{$\tilde{r}$}

{$(2)\, :a^2=1$
\includegraphics*[width=7cm]{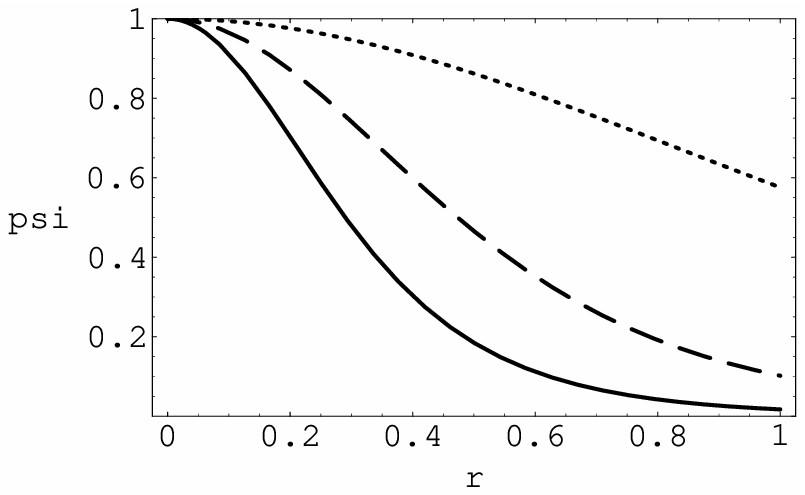}
\!\!\!\!\!\!\!\!\!\!\!\!\!\!\!\!\!\!\!\!\!\!\!\!\!\!\!\!\!\!\!\!\!\! %
$(3)\, :a^2=2$
\includegraphics*[width=7cm]{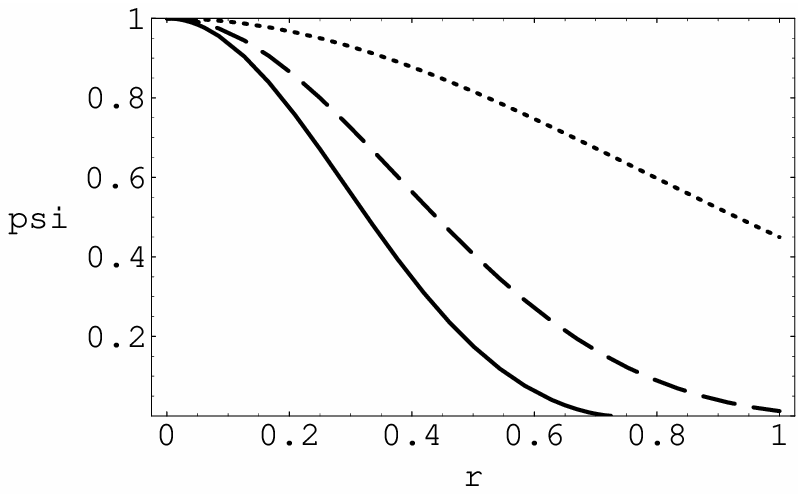}
\includegraphics*[width=7cm]{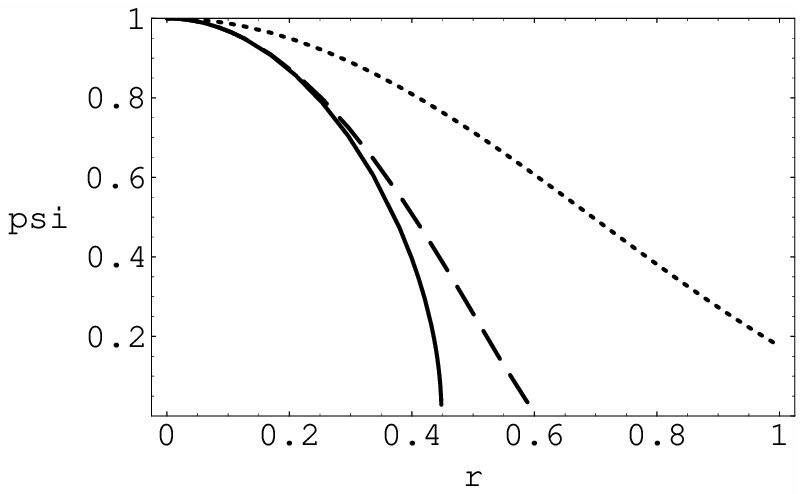}}
\caption{%
The density distribution of the isothermal sphere of ``extreme black holes" for different values of the coupling constant of the dilaton field; (1) $a^2=0$, (2) $a^2=1$ and (3) $a^2=2$. The solid line denotes $\delta=0$ (the high temperature limit), the broken line $\delta=1$ and the dotted line $\delta=10$.}
\label{fig:ebh}
\end{figure}

For solving explicitly eq.~(\ref{eq:263}), let us assume the spherical symmetry and define the following quantities:
\begin{eqnarray}
\label{eq:264}
n_0 &\equiv& e^{\beta\mu} \Bigl(\frac{m}{2\pi\beta}\Bigr)^{3/2} ,\\
\label{eq:265}
\tilde{r} &\equiv& \sqrt{G \,m \,n_0}\,r ,\\
\label{eq:266}
\delta &\equiv& 2\sqrt{3}\,G \,n_0 \beta ,
\end{eqnarray} 
where $G$ is the Newton constant. Then eq.~(\ref{eq:263}) for $V$ is
\begin{eqnarray}
\frac{1}{\tilde{r}^2}\Bigl(\tilde{r}^2 V(\tilde{r})^{'}\Bigr)'  %
    &=& 4\pi\,(1 + a^2)V(\tilde{r})^{\frac{3(3-a^2)}{2(1+a^2)}} %
               \Bigl(\, \cos\frac{\theta}{3} %
                + \frac{1}{\sqrt{3}}\sin\frac{\theta}{3}\,\Bigr)^{-3} ,
\label{eq:267}
\end{eqnarray}
where prime denotes derivative with respect to $\tilde{r}$, and
\begin{eqnarray}
\label{eq:268} 
 \theta(\tilde{r}) \:= \:\sin^{-1}\Bigl[ %
     (3-a^2) V(\tilde{r})^{\frac{(1-3a^2)}{2(1+a^2)}}\, \delta\Bigr] .
\end{eqnarray}

From eqs.~(\ref{eq:261}), (\ref{eq:267}) and (\ref{eq:268}), we have calculated numerically the relation between the radius $\tilde{r}$ and the particle density $|\psi|^2$ of the isothermal sphere of ``extreme black holes". (see Fig.~\ref{fig:ebh}.)  As seen from Fig.~\ref{fig:ebh}, we find that as $\delta$ is the smaller, that is, the temperature is the higher, the particles lump the more tightly, because the magnetic force acts among the particles as the attractive force in order $v^2$ from the electromagnetic field equation (\ref{eq:209}). And we see that as the coupling constant of the dilaton field is the larger, the particles lump the more tightly. In particular, there is a critical point whether an extent of the edge exists or not. For the case $a^2=3$, the effective lagrangian represents the free lagrangian, therefore the particles correspond to the free particles.


\section{Conclusion}

In this paper, we have derived the effective lagrangian of ``extreme black holes" in the low energy. For the finite temperature, we have obtained Hartree\,-\,Fock equation and then we have seen the structure of the isothermal sphere distribution of ``extreme black holes". In the high temperature, the gas of ``extreme black holes"  have been lumped  by the attractive force. We have seen that there is the critical point at $a^2\sim 1$ whether an extent of the edge exists or not.

\begin{figure}[h]
\centering
\psfrag{k1}{$\vect{k}$}
\psfrag{q1}{$\vect{q}$}
\psfrag{kq1}{$\vect{k}+\vect{q}$}
 \includegraphics*[width=5cm,height=2.5cm]{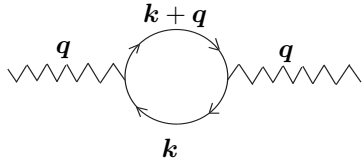}
\caption{Self-energy of the vector field}
\label{fig:vector}
\end{figure}

In future work, we will evaluate the correction by the self-energy of the vector field $\vect{A}$ (see Fig.~\ref{fig:vector}), consider the case of the low temperature, that is, the condensation, take the statistics of ``extreme black holes" into consideration and study the calculation by means of a lattice space.


\end{document}